# Electrical tunability due to coalescence of exceptional points in parity-time symmetric waveguides

Jin Wang,[1,2] Hui Yuan Dong,[3] Raymond P. H. Wu,[2,4] T. C. Mok,[2] and Kin Hung Fung[2,*]

[1]*Department of Physics, Southeast University, Nanjing 211189, China*
[2]*Department of Applied Physics, The Hong Kong Polytechnic University, Hong Kong, China*
[3]*School of Science, Nanjing University of Posts and Telecommunications, Nanjing 210003, China*
[4]*Department of Mathematics, The Hong Kong University of Science and Technology, Hong Kong, China*

We demonstrate theoretically the electric tunability due to coalescence of exceptional points in PT-symmetric waveguides bounded by imperfect conductive layers. Owing to the competition effect of multimode interaction, multiple exceptional points and PT phase transitions could be attained in such a simple system and their occurrences are strongly dependent on the boundary conductive layers. When the conductive layers become very thin, it is found that the oblique transmittance and reflectance of the same system can be tuned between zero and one by a small change in carrier density. The results may provide an effective method for fast tuning and modulation of optical signals through electrical gating.

PACS numbers:

A wide class of non-Hermitian Hamiltonians with parity-time (PT) symmetric complex potentials have been extensively studied since Bender et al. [1, 2] showed such systems can exhibit purely real spectra. By tuning the degree of non-Hermiticity, PT-symmetric systems may experience an abrupt phase transition between PT-symmetric phase with a real spectrum and PT-broken phase with a complex spectrum. The transition point is called the exceptional point (EP), at which two or more eigenvalues coincide. The existence of EPs have been investigated in various types of PT-symmetric physical systems [3–19], giving rise to a wide range of counterintuitive wave phenomena, such as loss-induced transparency [3], nonreciprocal Bloch oscillation [4], unidirectional invisibility [5–7], coexisting coherent perfect absorption and lasing [8, 9], enhanced spontaneous emission [10], and enhanced nano-particle sensing [11].

Recently, there has been strong interest in more complex phenomena closely related to the occurrence of EPs in multistate systems and high-order EPs [10, 20–27]. Multiple optical waveguide systems [20, 21] and photonic crystals [10, 22] have been proposed for the realization of high-order EPs. For example, Ding et al. [22] demonstrated theoretically two different kinds of EP evolution process in one-dimensional PT-symmetric photonic crystals. One type is the occurrence of a ring of EPs, leading to the restoration behavior of PT symmetry phase. Another type is the coalescence of EPs to form high-order singularity. Zhen et al. [23] observed experimentally that a ring of EPs could be supported near a Dirac-like cone in a two-dimensional photonic crystal slab. Also, the emergence and interaction of multiple EPs have been studied in a four-state acoustic system [24], exhibiting richer physical behaviors than those seen in two-state systems.

Taking the advantage of the abrupt change near an EP, one may tune the properties of a system drastically by a small change in certain parameters. However, previous studies on evolution process of EPs and various types of phase transitions, are mostly on tuning by geometrical parameters. In this Letter, we demonstrate that the PT-symmetric plasmonic waveguide bounded by imperfect conducting materials such as doped semiconductors (SC), may have electrically tunable emergence or coalescence of exceptional points which depend substantially on the free carrier density of the conducting boundaries. When the boundary layers become very thin, it is found that the sideway reflectance (and transmittance) of the same system can be fully tuned between zero and one by only a small change in plasma frequency, which is governed by the carrier density. Our results may provide an approach to modulate optical signals near the EPs in a controllable manner, which is significant to potential applications of non-Hermitian optical systems.

We start with a planar PT-symmetric plasmonic waveguide depicted in Fig. 1(a). Two dielectric lossy and gain slabs with identical thickness $d$ and balanced dielectric constants $\epsilon_L = \epsilon + i\tau$ and $\epsilon_G = \epsilon - i\tau$ are embedded in semi-infinite SC claddings. For the SC region, we consider a lossless Drude model to highlight the emergent features of PT potentials, with a permittivity $\epsilon_{SC} = \epsilon_\infty(1 - \omega_{p0}^2/\omega^2)$, where the plasma frequency of SC is $\omega_{p0} = \sqrt{ne^2/m^*\epsilon_\infty\epsilon_0}$, with $n$, $e$, $m^*$, $\epsilon_\infty$, and $\epsilon_0$ being the free carrier densities, the electron charge, the effective mass of free carrier, the high-frequency dielectric constant, and the vacuum permittivity, respectively. In this letter, we consider the propagation of TM-polarized light beams (i.e. only $E_x$, $E_y$, $H_z$ components of the electric and magnetic field are nonzero) along the $y$ direction and use the $e^{-i\omega t}$ time-dependent convention for oscillating fields.

To find the guided modes in the PT-symmetric plasmonic waveguides, we solve the Maxwell's equations in each region, yielding the field components, $H_z(x,y) = Ae^{\beta(x+d)+ik_y y}$ for $x < -d$ and $H_z(x,y) =$ $Be^{-\beta(x-d)+ik_y y}$ for $x > d$ in the SC boundary claddings, and $H_z(x,y) = [C\cos(k_{Lx}x) + D\sin(k_{Lx}x)]e^{ik_y y}$ for $0 < x < d$ and $H_z(x,y) = [E\cos(k_{Gx}x) + F\sin(k_{Gx}x)]e^{ik_y y}$ for $-d < x < 0$ in the core regions of two dielectric slabs. Here, $A$, $B$, $C$, $D$, $E$ and $F$ are the magnetic field amplitudes, $\beta$ denotes the positive decay parameters with $\beta = \sqrt{k_y^2 - \epsilon_{SC}\mu\omega^2/c^2}$ in order to guarantee the field exponential decay in the SC claddings, and $k_{Lx,Gx} = \sqrt{\epsilon_{Lx,Gx}\mu\omega^2/c^2 - k_y^2}$. With the requirement of continuity of magnetic field $H_z$ and its corresponding tangential electric field $E_y$ at each interface, we could obtain the dispersion relation for the guided modes given by

$$(\frac{k_{Gx}\epsilon_{SC}}{\epsilon_G\beta} - \frac{\epsilon_G\beta}{k_{Gx}\epsilon_{SC}})\tan[k_{Gx}d] + (\frac{k_{Lx}\epsilon_{SC}}{\epsilon_L\beta} - \frac{\epsilon_L\beta}{k_{Lx}\epsilon_{SC}})\tan[k_{Lx}d] + \tan[k_{Gx}d]\tan[k_{Lx}d](\frac{k_{Gx}\epsilon_L}{\epsilon_G k_{Lx}} + \frac{\epsilon_G k_{Lx}}{k_{Gx}\epsilon_L}) = 2.$$

(1)

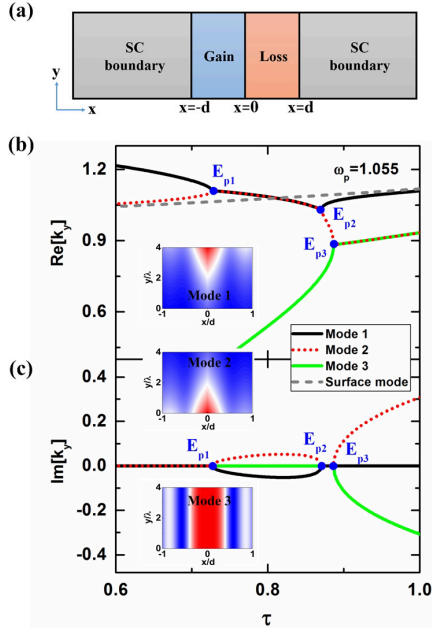

FIG. 1: (a) Schematic diagram of PT-symmetric plasmonic waveguides, containing semi-infinite SC cladding. (b)-(c) Real and imaginary parts of $k_y$ as a function of loss/gain strength $\tau$ when normalized plasma frequency $\omega_p = 1.055$. Three guided modes are shown by solid (black), dotted (red), and solid (green) lines, respectively. The surface mode at the interface between semi-infinite lossy and gain materials is also shown by dotted (gray) lines. Possible EPs are marked with $E_{p1}$, $E_{p2}$ and $E_{p3}$, respectively. Inset of Fig. 1(b) plots the moduli of field component $E_y$ for Modes 1,2, and 3 at $\tau = 0.8$.

Eigenmodes of the waveguides were found by using a commercial root-solver (Mathematica FindRoot), to search for numerical roots $k_y$ of Eq. (1) at a fixed $\omega$. Two types of mode solutions could be identified by the imaginary part of corresponding complex propagation constant $k_y$: purely real modes with $Im[k_y] = 0$ and a pair of complex-conjugate modes, decaying and growing modes, with $Im[k_y] \neq 0$. These two possibilities correspond to PT-symmetric and PT-broken regimes.

In the following calculation, we fix the real part of lossy or gain material with $\epsilon = 2$, and vary the imaginary part $\tau$. For SC materials, we set $\epsilon_\infty = 15.68$ [28], and tune its plasma frequency electrically by varying the free-carrier densities, i.e. by applying a gate voltage. Without loss of generality, we focus on the regime with a given working frequency $\omega = 2\pi c/\lambda_0$, where $\lambda_0$ is the free-space wavelength, and set the thickness of each slab at $d = 0.58\lambda_0$. Figures 1(b) - 1(c) show the propagation constant $k_y$ as a function of loss/gain strength $\tau$ for a particular plasmon frequency of SC, i.e. $\omega_p = 1.055$, where $\omega_p$ is defined as the normalized plasma frequency $\omega_p \equiv \omega_{p0}/\omega$ for simplicity. In this configuration, it is found that three possible exceptional points, marked by $E_{p1}$, $E_{p2}$ and $E_{p3}$, occur at $\tau = \tau_1 = 0.726$, $\tau = \tau_2 = 0.87$ and $\tau = \tau_3 = 0.886$, respectively. Two guided modes are supported as Mode 1 and Mode 2 at the initial condition of $\tau = 0$. By comparing with the single interface surface plasmon mode when there is no loss or gain (dashed line in Fig. 1), we see that Mode 1 and Mode 2 are originally from the two coupled surface plasmon modes support near the inner surfaces of the conducting boundary layers. At small $\tau(\tau < \tau_1)$, $k_y$ remains purely real for both of them, and the usual mode symmetry holds. When $\tau$ approaches $\tau_1$, the first exceptional value $E_{p1}$ occurs, Modes 1 and 2 merge into a pair of complex modes with complex-conjugate $k_y$, and the original mode symmetry is broken. Modes propagation will be either amplified or attenuated. Note that such PT-phase transition corresponding to first EP ($E_{p1}$) could take place for both TE and TM polarizations with either SC or dielectric cladding [29]. Surprisingly, further increase of gain/loss strength results in restoration of PT symmetry phase beyond a second EP ($E_{p2}$) and purely two real modes reappear at $\tau_2 < \tau < \tau_3$. Notice that another purely real mode, named as Mode 3, could be supported when non-Hermiticity parameter is turned on, and finally merge with Mode 2 into a pair of complex modes at the third EP ($E_{p3}$). The inset field plots in Figs. 1(b) and 1(c) show that Mode 3 is a propagating mode centered between the gain and loss media when $\tau = 0.8$. The insets also show that Modes 1 and 2 are mostly localized at the center of the waveguide when $\tau = 0.8$, which is near the coalescence of exceptional points. We will show that such a possible coalescence and re-bifurcation is a result of multiple state interaction to be discussed below, which has also been found in the other multistate systems, such as 1D PT-symmetric photonic crystal systems [22] and coupled acoustic cavities [24].

To show the criticality, we demonstrate two other cases with different plasma frequencies. For lower plasma fre-quency, i.e. $\omega_p = 1.048$ shown in Fig. 2(a)-(b), we see that purely real Mode 1 and Mode 2 mix and merge at $E_{p1}$, bi-exceptional points $E_{p2}$ and $E_{p3}$ coalesce with each other and disappear, and no re-entry behavior of PT symmetry could happen with further increase of loss/gain strength, even when purely real Mode 3 meets the complex modes. At this stage, these modes can be regarded as nearly independent, even though the coupling effect does exist. However, the situation could be changed by tuning to a higher $\omega_p$, i.e. $\omega_p = 1.062$ seen in Fig. 2(c)-(d). The system remains in PT-symmetric phase until Modes 2 and 3 couple and combine together when the gain/loss strength approaches the only exceptional value $\tau_3 = 0.878$. The value is close to that appeared in Fig. 1. No restoration of PT symmetry could also be seen. For this case, the dominant mixing mechanism between Modes 2 and 3 tends to pull Mode 1 out of the PT-broken phase. Therefore, in our proposed system, we could possibly achieve the multiple EPs and various phase transition behaviors by tuning flexibly the coupling between different guided modes dependence of the plasma frequency of SC.

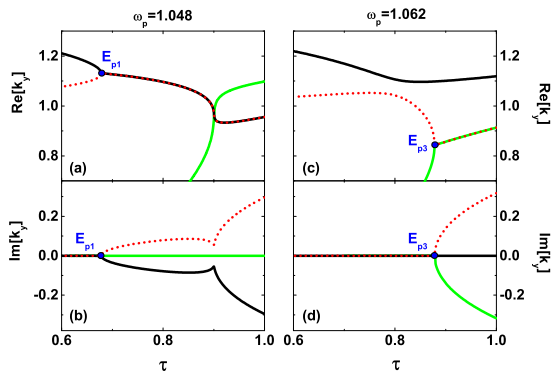

FIG. 2: Real and imaginary parts of $k_y$ as a function of loss/gain strength $\tau$ for normalized plasma frequency $\omega_p = 1.048$ [(a)-(b)] and 1.062 [(c)-(d)]. The occurrence of EPs are also indicated. Other parameters are the same as Fig. 1.

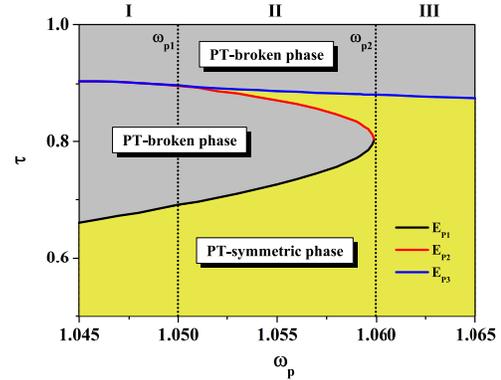

FIG. 3: Trajectories of the exceptional points in the $(\omega_p, \tau)$ space: $E_{p1}$ (black line), $E_{p2}$ (red line), and $E_{p3}$ (blue line). The gray region stands for the PT-broken phase, and the yellow region for PT-symmetric phase. Three different regimes, class I, II and III, are separated by the vertical dotted lines, which are labelled with $\omega_{p1}$ and $\omega_{p2}$, respectively.

We summarize the trajectories of exceptional points, $E_{p1}$, $E_{p2}$, and $E_{p3}$ in the $(\omega_p, \tau)$ space in Fig. 3. Three different regimes (class I, II and III) could be divided by the coalescence and bifurcation of EPs, substantially dependent on the competition interacting effect among Modes 1, 2, and 3. At lower plasma frequency [$\omega_p < \omega_{p1}(\approx 1.050)$], in class I of the system, the strong coupling effect between Modes 1 and 2 could be seen, leaving Mode 3 remain purely real. Only single phase transition appears at $E_{p1}$ from PT-symmetric phase to PT-broken phase and the other two EPs, $E_{p2}$ and $E_{p3}$ coalesced with each other in this regime. As the system enters class II [$\omega_{p1} < \omega_p < \omega_{p2}(\approx 1.060)$], the attractive interaction between Modes 2 and 3 increases gradually, leading to the splitting of $E_{p2}$ and $E_{p3}$, and the emergence of the re-entry behavior between $E_{p2}$ and $E_{p3}$. Moreover, as $\omega_p$ increases further, it will make this dominant interaction stronger, and $E_{p1}$ and $E_{p2}$ start to move towards each other and finally merge at $\omega_{p2}$. Therefore, in this regime multiple EPs are found and various phase transitions could happen between PT-symmetric phase and PT-broken phase. Eventually, in class III [$\omega_p > \omega_{p2}$], the mixing mechanism between Modes 2 and 3 governs the entire system, and pull Mode 1 out of PT-broken phase. There exists only single phase transition at the critical point $E_{p3}$ in class III.

The dispersion relations and PT phase transition can also be revealed through scattering behavior if the thickness of the SC boundary material in Fig. 1(a) is reduced to a finite thickness $w$. Figure 4 plots the reflectance $R_L$ in the left and $R_R$ in the right of the same structure, embedded in a uniform surrounding medium $\epsilon = 2$, when the plasma frequency is tuned to $\omega_p = 1.055$ and 1.062, respectively. The reflection amplitude displays dips when the system is in PT-symmetric phase, while it stays high in PT-broken phase. It is found that small tuning with $\omega_p$ could possibly induce a large variation on reflection. Note that either left or right reflectance dips in each case are closely related to its associated purely real band curves for the infinite systems shown by black dashed lines.

Furthermore, we demonstrate the effective tuning of transmittance or reflectance for the case of thin boundary layers. Fig. 5(a) shows the transmittance T and reflectance $R_L$, $R_R$ as the function of $\omega_p$ for a particular case of $\tau = 0.8$ and incident angle of light illumination $\theta = 46.7^o$. It is found that two transmission resonances from left and right occur at $\omega_p = 1.062$ and 1.0628, respectively. Note that the general conservation rule $|1 - T| = \sqrt{R_L R_R}$ [30] should be obeyed in this 1D PT-symmetric system. The results are verified by 2D finite-element simulations using Comsol Multiphysics. Figure 5(b)-(c) depict the spatial field distribution calculated using COMSOL for left incidence at a particular incident angle $\theta = 46.7^o$ with $\omega_p = 1.055$ and 1.062, respectively. Strong reflectance occurs due to the suppression of transmission resonance for the case of $\omega_p = 1.055$, whereas by tuning slightly to higher plasma frequency $\omega_p = 1.062$, the left transmission resonance is supported due to the excitation of Mode 2 [as seen from Fig. 2(c)(d)], which is closely related to the surface plasmon-polariton mode, and nearly full transmission for left incidence is thereby achieved.

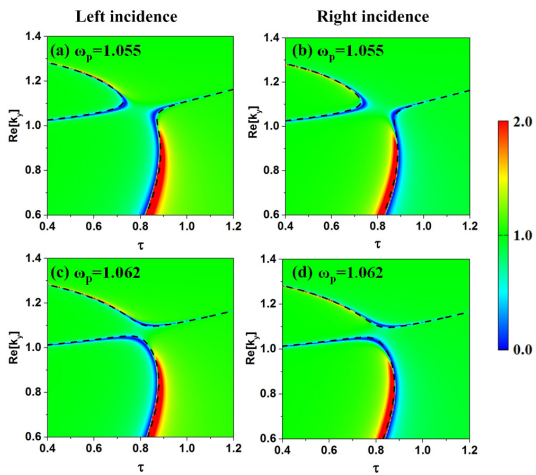

FIG. 4: Reflectance of finite-size PT-symmetric slabs embedded in a uniform surrounding medium $\epsilon = 2$. (a) Light incident from left, $\omega_p = 1.055$. (b) Light incident from right, $\omega_p = 1.055$. (c) Light incident from left, $\omega_p = 1.062$. (d) Light incident from right, $\omega_p = 1.062$. For this finite-size system, we truncate the semi-infinite SC cladding depicted in Fig. 1(a) into two thin slabs with its identical thickness $w = 0.2\lambda_0$. For comparison, the associated purely real bands for the infinite systems are also shown with black dashed lines.

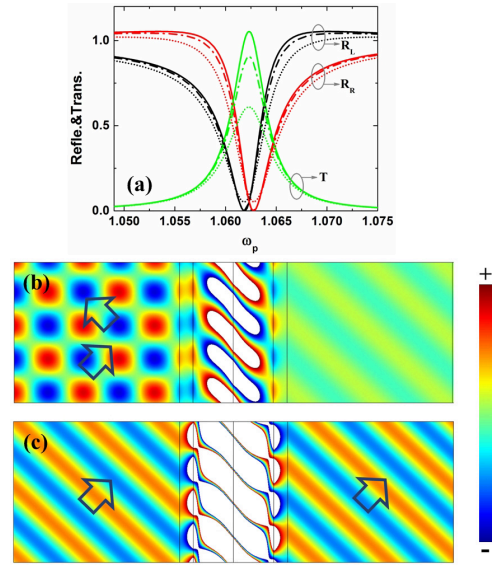

FIG. 5: (a) Transmittance $T$, left and right reflectance $R_L$, $R_R$ as a function of normalized plasma frequency $\omega_p$ for $\tau = 0.8$ and incident angle of light illumination $\theta = 46.7^o$. Three regimes with different loss $\epsilon'$ in SC layers as shown by solid lines [$\epsilon' = 0$], dash-dotted [$\epsilon' = 0.005$] and dotted lines [$\epsilon' = 0.02$], respectively. (b)(c) The associated field distribution under left illumination at (b) $\omega_p = 1.055$ and (c) $\omega_p = 1.062$, when the loss in SC layers is absent.

We emphasize that such a variation in the evolution process of three different regimes summarized above depends on the plasma frequency of SC, which could be controlled by active voltage using a carrier density tuning mechanism [31]. Such nearly perfect tuning in Fig. 5(b) could only be achieved under certain assumptions. First, it is assumed that the penetration depth of the field has a large overlap with the depletion region where the charges concentrate. Second, the average percentage of increase in charge careers has to be larger than 2% such that the plasma frequency has a change of about 1%. Finally, owing to the inevitable loss in the SC layers, we see how it affects the tuning performance of transmittance or reflectance, also shown in Fig. 5(a). As the loss $\epsilon'$ in SC layers is increased from $\epsilon' = 0.005$ to $\epsilon' = 0.02$, the left and right transmission resonance may be suppressed partly and small reflection could be seen. Although the required level of gain could be a potential issue to be solved, it could be lowered by tuning the geometrical parameters, the working frequency, or index contrast in our waveguide heterostructures.

In conclusion, we show that the emergence and coalescence of multiple exceptional points and phase transitions could be tuned electrically in a PT-symmetric waveguide. Such a flexible tuning effect is a result of multiple state interaction, possibly modulated by the free carrier densities of conductive boundary layers. Meanwhile, full transmission or strong reflection could be achieved through such a finite-size PT-symmetric system by application of the abrupt changes due to the emergence and coalescence of multiple exceptional points. Our results may suggest an effective method for fast tuning and modulation of optical signals through electrical gating.

**Funding**. Hong Kong Research Grant Council (AoE/P-02/12, 15300315); The Hong Kong Polytechnic University (G-UC70, G-YBPT); Natural Science Foundation of Jiangsu Province (BK20160878); National Natural Science Foundation of China (NSFC) (11204036); Nanjing University of Posts and Telecommunications (NY215090).

66